\documentclass[aps,prl,twocolumn,superscriptaddress]{revtex4}

\usepackage{graphicx}
\usepackage{amsmath}
\usepackage{color}

\begin{document}

\title{Circular conversion dichroism in planar chiral metamaterials}

\author{V. A. Fedotov}
\affiliation{EPSRC Nanophotonics Portfolio Centre, School of Physics
and Astronomy, University of Southampton, SO17 1BJ, UK}

\author{P. L. Mladyonov}
\affiliation{Institute of Radio Astronomy, National Academy of
Sciences of Ukraine, Kharkov, 61002, Ukraine}

\author{S. L. Prosvirnin}
\affiliation{Institute of Radio Astronomy, National Academy of
Sciences of Ukraine, Kharkov, 61002, Ukraine}

\author{A. V. Rogacheva}
\affiliation{EPSRC Nanophotonics Portfolio Centre, School of Physics
and Astronomy, University of Southampton, SO17 1BJ, UK}

\author{Y. Chen}
\affiliation{Central Microstructure Facility, Engineering and
Instrumentation Department, Rutherford Appleton Laboratory,
Oxfordshire, OX11 0QX, UK}

\author{N. I. Zheludev}
\email{n.i.zheludev@soton.ac.uk}
\homepage{www.nanophotonics.org.uk}
\affiliation{EPSRC Nanophotonics Portfolio Centre, School of
Physics and Astronomy, University of Southampton, SO17 1BJ, UK}

\date{\today}

\begin{abstract}
We report the first experiential observation and theoretical analysis
of the new phenomenon of \emph{planar chiral circular conversion
dichroism}, which in some aspects resembles the Faraday effect in
magnetized media, but does not require the presence of a magnetic
field for its observation. It results from the interaction of an
electromagnetic wave with a planar chiral structure patterned on the
sub-wavelength scale, and manifests itself in asymmetric transmission
of circularly polarized waves in the opposite directions through the
structure and elliptically polarized eigenstates. The new effect is
radically different from conventional gyrotropy of three-dimensional
chiral media.
\end{abstract}

\maketitle

Since Hetch and Barron \cite{Hetch} and Arnaut and Davis
\cite{Arnaut} first introduced planar chiral structures to
electromagnetic research they have become the subject of intense
theoretical \cite{Prosvirnin,Bedeaux} and experimental investigations
with respect to the polarization properties of scattered fields
\cite{Papakostas1,Papakostas2,OptComm}. It was understood by many
that planar chirality is essentially different in symmetry from
three-dimensional chirality. Whereas in three-dimensional chiral
structures the sense of perceived rotation remains unchanged for
opposing directions of observation (think, for example, of a helix
observed along its axis), planar chiral structures possess a sense of
twist that is reversed when they are observed from opposite sides of
the plane to which the structure belongs. Consequently, if planar
chiral structures were to exhibit a polarization effect (due to this
twist) for light incident normal to the plane, the sense of the
effect would be reversed for light propagating in opposite
directions. Such behavior has never been observed before, but if
proven would be of profound benefit to the development of a new class
of microwave and optical devices.

In this paper we report such a polarization sensitive effect. It is a
previously unknown fundamental phenomenon of electromagnetism that
asymmetric materials can generate behaviors that in some ways
resemble the famous non-reciprocity of the Faraday effect, which
emerges when a wave propagates through a magnetized medium. However,
the phenomenon reported here does not require the presence of a
magnetic field and results from an electromagnetic wave's
transmission through a chiral planar structure patterned on the
sub-wavelength scale. Both in the Faraday effect and in that produced
by planar chirality, the transmission and retardation of a circularly
polarized wave are different in opposite directions. In both cases
the polarization eigenstates, i.e. polarization states conserved on
propagation, are elliptical (circular).

\begin{figure}
\includegraphics[width=80mm]{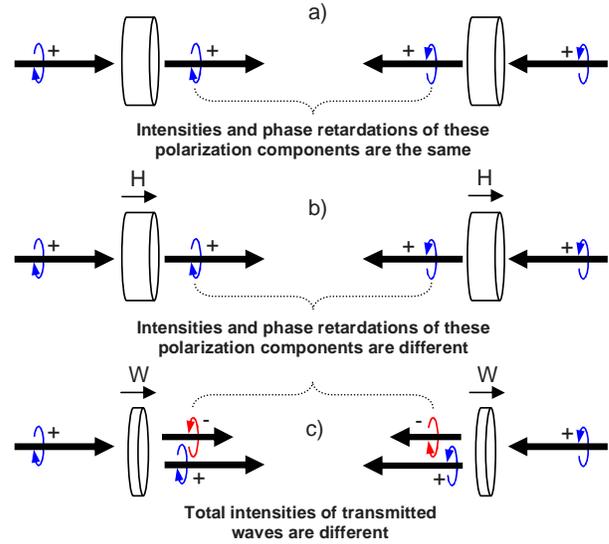}
\caption{Propagation of circularly polarized waves through (a) three
dimensional chiral medium (b) Faraday and (c) planar chiral media.
The colored circular arrows denote circular polarization states: blue
for RCP, red for LCP. Propagation through an anisotropic planar
chiral structure leads to polarization conversion. The vector
$\mathbf H$ denotes an external magnetic field, the vector $\mathbf
W$ indicates the twist of the planar chiral structure.}
\end{figure}

There are also essential differences between the two phenomena. The
asymmetry of the Faraday effect with respect to propagation in
opposite directions applies to the transmission and retardation of
the \emph{incident} circularly polarized wave itself. The planar
chirality effect leads to the (partial) conversion of the incident
wave into one of opposite handedness, and it is the efficiency of
this \emph{conversion} that is asymmetric for the opposite directions
of propagation (see Fig.1). The eigenstates of an anisotropic Faraday
medium are two elliptically polarized waves of \emph{opposite}
handedness. The eigenstates of a planar chiral medium are two
elliptical polarizations of the \emph{same} handedness. The newly
observed effect of planar chirality is also radically different from
conventional gyrotropy in three-dimensional chiral media, such as the
sugar solution which is complectly symmetric with respect to the
inversion of the direction of propagation. The later property of
conventional gyrotropy results from the fact that the sense of
helicity of a spiral (or any three-dimensional chiral object for this
matter) does not depend on the direction at which it is observed.
The propagation asymmetry observed in planar chiral media is
nevertheless compatible with the general notion of reciprocity as
defined by the Lorentz Lemma. Moreover, the asymmetric effect in
planar chiral structures is inherently linked to losses in the
structure. This will be explained in more detail below.

If a medium is described by a complex transmission matrix $\chi$ for
the field amplitudes of the incident $E^0$ and transmitted
 $E^T$ circularly polarized waves, then $E^T_{i} = \chi_{ij}E^0_j$, where the indices $i$ and $j$
correspond to the transmitted and incident polarization states, which
could be either right (RCP, $+$) or left circular polarizations (LCP,
$-$). The matrix $\chi^F$ for an isotropic magnetized Faraday medium
and matrix $\chi^{3D}$ for isotropic three-dimensional chiral medium
are  \emph{diagonal} matrices. In the Faraday medium the transmission
matrices for opposing directions of propagation (denoted by arrows),
i.e. $\overrightarrow{\chi^F}$ and $\overleftarrow{\chi^F}$, are
related by the permutation of their diagonal elements, while in the
three-dimensional chiral medium the transmission matrices for
opposing directions of propagation, i.e. $\overrightarrow{\chi^{3D}}$
and $\overleftarrow{\chi^{3D}}$, are identical. In these terms the
transmission matrix for a planar chiral anisotropic medium is a
\emph{non-Hermitian} matrix with equal diagonal elements $\chi^{2D}=
\left\{\begin{array}{ll}
    \alpha & \beta \\
    \gamma & \alpha
\end{array}\right\}$. The transmission matrices for opposing directions of propagation will be
mutually transposed: $\overrightarrow{\chi^{2D}_{ij}}$ =
$\overleftarrow{\chi^{2D}_{ji}}$. For a given direction of
propagation the equality of the diagonal elements
$\chi_{++}=\chi_{--}=\alpha$ implies that losses and retardation are
identical for RCP and LCP waves passing through the structure.
However, since $\chi_{+-}\neq \chi_{-+}$, switching between RCP to
LCP incident waves leads to a change in the intensity and phase of
the corresponding LCP and RCP converted circular components.

The new propagation phenomenon described by the matrix $\chi^{2D}$
has been observed in a chiral `fish-scale' planar structure. This is
a chiral version of a new type of electromagnetic meta-material, the
non-chiral form of which was recently investigated for its frequency
selective properties and `magnetic wall' behavior \cite{FS}. The
chiral `fish-scale' is a 2-dimensional continuous pattern of tilted
meanders existing in two enantiomeric (mirror) forms interconverted
by reflection across a line in the plane of the structure (see
Fig.2). On the basis of both intuitive perception and strict
\begin{figure}
\includegraphics[width=80mm]{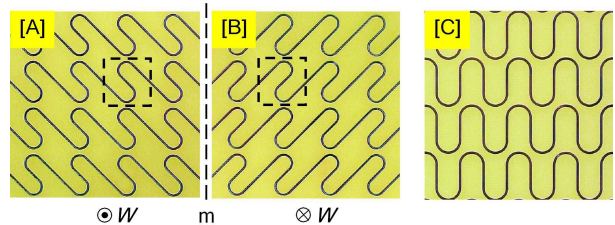}
\caption{Fragments of the two enantiomeric forms A and B of the
chiral `fish-scale' pattern of copper strips on a dielectric
substrate, for which planar chiral circular conversion dichroism was
observed at normal incidence. The dashed line box indicates the
elementary translational cell of the structure. Part C shows a
non-chiral fish-scale structure, which exhibited no chiral effect in
transmission.}
\end{figure}
mathematical definition \cite{ChiralTheory}, the left-tilted
fish-scale A shows an overall anti-clockwise twist, while the
right-tilted fish-scale B twists clockwise. An axial `twist' vector
$\mathbf W$, governed by the cork-screw law, may be associated with
each of the enantiomers.

In the experiments reported here we used a metallic fish-scale
structure, with a $15 \times 15$ $mm$ square translation cell, etched
from a $35$ $\mu m$ copper film on a $1.5$ $mm$ thick dielectric PCB
material substrate (see Fig.~1). The width of the strips was $0.8$
$mm$. The overall size of the samples was approximately $220 \times
220$ $mm$. We studied the reflection and transmission of this
structure in the $4-18$ $GHz$ spectral range, performing measurements
in an anechoic chamber using a vector network analyzer (Agilent,
model E8364B). The structure does not diffract electromagnetic
radiation at frequencies lower than $20$ $GHz$ so a single-wave
regime was achieved in the experiments. We investigated transmission
through the two enantiomeric forms of the chiral fish-scale structure
at normal incidence. The handedness of the pattern depends on whether
it is observed from one side of the screen or the another. In this
paper the handedness will be defined as seen by the \emph{incident}
electromagnetic wave and the structure will be referred to as being
`clockwise' if the vector $\textbf{W}$ is \emph{antiparallel} to the
direction of the incident wave. All parameters for the incident and
transmitted waves will be defined in a right-handed cartesian
coordinate frame where the $Z$ axis is directed perpendicular to the
plane of the structure along the direction of the incident wave,
while the $X$ and $Y$ axes are directed along and perpendicular to
the meander lines respectively.

Measuring the circular polarization transmission matrix $\chi^{2D}$
directly requires circularly polarized emitters and receivers.
Although circular polarization antennae exist, they are only capable
of producing high purity circular polarization in a narrow spectral
range and are therefore not suitable for broadband measurements.
Instead, we used high-quality broadband linearly polarized and
log-periodic antennas (Schwarzbeck~M.~E. model STLP 9148) and
measured the complex transmission matrix $t$ for linearly polarized
fields. The complex circular polarization transmission matrix was
then calculated as follows:
\\
\\
$\overrightarrow{\chi^{2D}} = \frac{1}{2}\left\{
\begin{array}{ll}
    {t_{xx}+t_{yy}+i(t_{{xy}}-t_{{yx}})} &  ~{t_{xx}-t_{yy}-i(t_{{xy}}+t_{{yx}})}\\
    {t_{xx}-t_{yy}+i(t_{{xy}}+t_{{yx}})} &  ~{t_{xx}+t_{yy}-i(t_{{xy}}-t_{{yx}})}
\end{array}
\right\}$\\

In the following we also use an intensity transmission and conversion
matrix $\Xi_{ij}$ =$|\chi_{ij}|^2$. The matrix $\Xi$ for fish-scale
structures A and B, where transmission is measured for waves entering
the structure from the side of the metal pattern, will be denoted
$\overrightarrow{\Xi^{A}}$ and $\overrightarrow{\Xi^{B}}$
respectively. Measurements were also conducted for waves entering the
structures from the opposite side, i.e. through the dielectric layer
first. The corresponding transmission matrices are
$\overleftarrow{\Xi^{A}}$ and $\overleftarrow{\Xi^{B}}$.

The transmission properties of the fish-scale structure were also
rigorously modelled using the method of moments described in
\cite{MethodMoments}. This widely used computational approach is
based on a vectorial integral equation for the surface current
induced by the electromagnetic wave on the metal strips of the
structure, which are assumed to form an infinitely thin perfectly
conducting patterned with topography identical to the structures used
in the experiment (see Fig.~2). This analysis takes into account
electromagnetic losses in the dielectric substrate supporting the
metal structure, which are introduced through the imaginary part of
the complex relative permittivity of the substrate material
$\varepsilon= 4.5+i0.2$. The results of these calculations show
remarkable agreement with those of the experiments. They are
presented in figures 3 and 4 as continuous lines alongside the
experimental points.

\begin{figure}
\includegraphics[width=80mm]{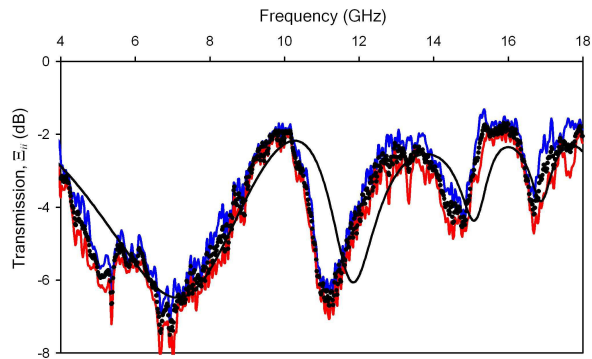}
\caption{Mean value of $\Xi_{ii}$ measured for both enantiomeric
forms (A and B) of the structure, both incident polarization states
(RCP and LCP) and in both forward and backward directions of
propagation. The mean value, an average over eight different
combinations of sample and wave handedness and propagation direction,
is represented by the black circles. The solid red and blue lines
show the limits of the variation in transmission for the four
configurations. The small variation in the results illustrates that
all eight experiments yield the same value of $\Xi_{ii}$ for given
frequency. The solid black line shows the theoretical value of
$\Xi_{ii}$ calculated using the method of moments. Within the model
$\Xi_{ii}$ does not depend on the direction of propagation and
handedness of the structure or polarization state. Therefore the
structure shows no conventional optical activity or
Faraday-effect-like polarization effect.}
\end{figure}

The data presented in Fig.~3 illustrates that within experiential
accuracy, the diagonal elements of matrix $\chi^{2D}$ are equal
across the whole spectral range of interest. Moreover, the diagonal
elements are the same for the two enantiomeric fish-scale structures
A and B and for propagation in both directions. In Fig.~3 we present
the squared modulus of the diagonal elements, but their arguments
linked to the phase retardation of the corresponding circular
components of the transmitted waves are also equal within
experiential accuracy. The fact that the diagonal elements of the
transmission matrix are independent of the propagation direction and
the chirality of the sample is fully consistent with the outcome of
calculations based on the method of moments. This indicates that the
structure does not manifest a polarization effect of the same
symmetry as conventional optical activity or the optical Faraday
effect in bulk media. However, it shows an intriguing new \emph{
assymetric polarization conversion} effect.

The data presented in Fig.~4 illustrates polarization conversion and
the asymmetric properties of the chiral fish-scale structures.
Fig.~4a shows the phase difference between converted polarization
components resulting from excitation with circularly polarized waves
of opposite handedness, measured as $\Delta\phi =Arg \{\chi_{-+}\}-
Arg \{\chi_{+-}\}$. Here one can see that the differential phase
delays for the two enantomeric forms of the structure have opposite
signs, i.e. $\Delta \phi^A = - \Delta \phi^B$. On the other hand, for
a given structure the differential phase delay observed in the
forward direction is the same as the delay for its \emph{enantiomer}
observed in the \emph{opposite} direction, i.e.
$\overrightarrow{\Delta \phi^A} = \overleftarrow{\Delta \phi^B}$.
Similar symmetries are observed for the normalized difference in the
intensities of the converted waves, measured as $\Delta \Xi = 2
(\Xi_{-+ } -\Xi_{+-}) / (\Xi_{-+} + \Xi_{+-})$. Once again
$\overrightarrow{\Delta \Xi^A} = - \overrightarrow{\Delta \Xi^B}$ and
$\overrightarrow{\Delta \Xi^A} = \overleftarrow{\Delta \Xi^B}$. This
indicates that the perceived sense of rotation of the planar chiral
structure (which depends on whether the structure is observed from
one side or another) controls the sing of the effect. In other words
it matters if the wave vector of the incident wave is parallel or
anti-parallel to the vector of structure's twist $\mathbf W$ (see
Fig. ~1). This is in sharp contrast with conventional gyrotropy in
three-dimensional chiral media where observable effect do not depend
on the direction of propagation.

It should be noted that our raw experiential data for the cartesian
transmission matrix, and the results of computational analysis, show
that in all cases $\overrightarrow{t_{xy}}=\overrightarrow{t_{yx}}$,
$\overleftarrow{t_{xy}}=\overleftarrow{t_{yx}}$ and
$\overrightarrow{t_{xy}}=\overleftarrow{t_{yx}}$,
$\overleftarrow{t_{xy}}=\overrightarrow{t_{yx}}$. The last two of
these equalities constitute the requirement imposed on transmission
through a planar chiral structure by the Lorentz Lemma
\cite{Lorentz}. In terms of circular polarizations the Lemma may be
re-written as an equality $\overrightarrow{\chi^{2D}_{ij}}$ =
$\overleftarrow{\chi^{2D}_{ji}}$, which holds within experimental and
computational tolerance as illustrated in Fig.~4. Thus, the effect
presented here does not mount any challenge to the validity of the
reciprocity Lemma. The somewhat surprising compatibility of this
asymmetric effect with the Lorentz Lemma is easily explained by the
fact that the asymmetry observed here results from the asymmetry of
polarization conversion for circular polarizations in the opposite
directions of propagation, i.e. $\overrightarrow{\chi^{2D}_{-+}} \neq
\overleftarrow{\chi^{2D}_{-+}}$, while the Lorentz Lemma only
requires that $\overrightarrow{\chi^{2D}_{-+}} =
\overleftarrow{\chi^{2D}_{+-}}$. In comparison, for the Faraday
effect $\overrightarrow{\chi^{F}_{++}} \neq
\overleftarrow{\chi^{F}_{++}}$ and the Lorentz Lemma does not hold,
nor indeed is it supposed to in the presence of a magnetic field.

\begin{figure}
\includegraphics[width=80mm]{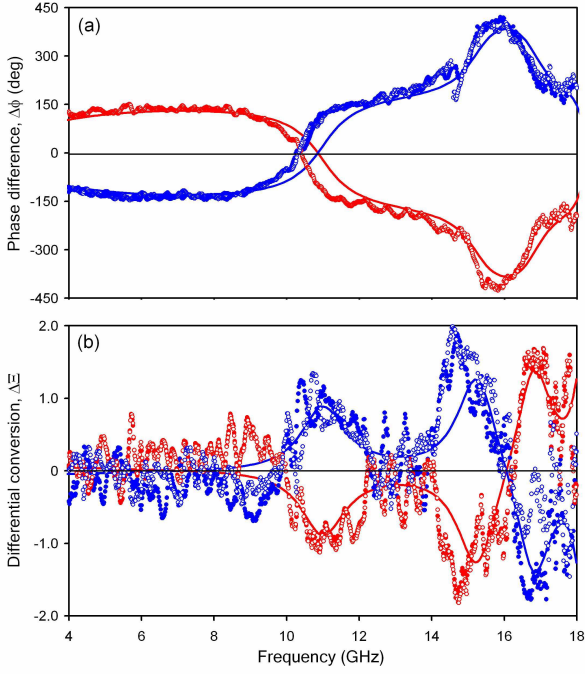}
\caption{Planar chiral circular conversion dichroism in chiral
fish-scale meta-material: polarization conversion in the two
enantiomers of the planar chiral structure in the forward and
backward directions of propagation. The upper graph shows the phase
difference $\Delta\phi$ between converted polarization components
resulting from excitation with circularly polarized waves of opposite
handedness. The lower graph shows the normalized differential
intensity of conversion, $\Delta \Xi$. The blue and red filled
circles correspond respectively to the A and B enantiomers for the
forward direction, while the blue and red empty circles correspond to
B and A forms for the reverse direction. Note, that similar results
in the \emph{opposite} directions are obtained for the
\emph{enantiomeric }forms of the structure.  The solid curves are
obtained using method of moments.}
\end{figure}

The origin of the assymetric interaction may be traced to the
structure of the transmission matrix $\chi^{2D}$. It may be presented
as a sum $\chi^{2D}= \chi^{2D}_0
+it_{xy}\mbox{sign}(\mathbf{kW})\widehat{g}$, where
\\
\\
$\chi^{2D}_0= \frac{1}{2}\left\{
\begin{array}{ll}
    {t_{xx}+t_{yy}} &  ~{t_{xx}-t_{yy}}\\
    {t_{xx}-t_{yy}} &  ~{t_{xx}+t_{yy}}
\end{array}
\right\}$\\
and is a symmetric matrix and $\widehat{g} = \left\{
\begin{array}{lr}
    0 &  -1\\
    1 &  0
\end{array}
\right\}$ is an anti-symmetric matrix. Here $\mathbf{k}$ is the wave
vector of the incident wave. The antisymmetric part is proportional
to a pseudo-scalar combination $\mathbf{kW}$ that changes its sign on
reversal of the direction of propagation. It gives rise to the
difference between $\overrightarrow{\chi^{2D}}$ and
$\overleftarrow{\chi^{2D}}$ and is therefore responsible for the
direction-dependent transmission. Essentially, the anti-symmetric
part is also proportional to $t_{xy}$, i.e. it may only exist in
anisotropic patterns of low symmetry (if the structure possesses a
4-fold symmetry axis, $t_{xy}=0$). Moreover, it is only in
dissipative systems that $t_{xy}$ cannot be eliminated by the choice
of an appropriate coordinate system. In other words, transmission
asymmetry is only possible in anisotropic dissipative planar chiral
structures.

The origins of $\Delta\phi$ and $\Delta\Xi$ are completely different.
$\Delta\phi$ results from the structures' anisotropy. Its value
depends on how enatiomeric samples are mutually oriented in the
experiment. More specifically it depends on the orientation of the
meander's direction of both enantiomers with respect to the line of
mirror symmetry inter-converting them. $\Delta\phi$ does not lead to
any asymmetrical effects that can be detected in the intensity of
propagating waves. In contrast, $\Delta \Xi$ results in an
\emph{observable intensity effect}. The effect is astonishing: the
planar chiral structure is more transparent to a circularly polarized
wave from one side than from another. For instance for an incident
RCP wave the total transmitted intensity in the forward direction is
given by $\overrightarrow{\Xi^A_{++}}+ \overrightarrow{\Xi^A_{-+}}$,
while in the opposite direction it is given by
$\overleftarrow{\Xi^A_{++}}+ \overleftarrow{\Xi^A_{-+}}$. In spite of
the fact that $\overrightarrow{\Xi^A_{++}} =
\overleftarrow{\Xi^A_{++}}$, polarization conversion is asymmetric,
i.e. $\overrightarrow{\Xi^A_{-+}} \neq \overleftarrow{\Xi^A_{-+}}$,
leading to a difference in the total transmitted intensities. No such
dependence on the direction of propagation would be seen in a
lossless chiral structure or an anisotropic non-chiral planar or bulk
material of any description (indeed, no $\Delta \Xi$ was detected for
the anisotropic non-chiral fish-scale structure C). Therefore this
phenomenon is somewhat analogous to the magnetic circular dichroism
effect (MCD) and may be called \emph{planar chiral circular
conversion dichroism} (PCD). Our calculations show that it is
inherently linked to losses and increases in proportion to the
imaginary part of the complex relative permittivity of the substrate
material. Lossless planar chiral metamaterial shall not display
asymmetric transmission for a circularly polarized wave. As we
already mentioned above, this asymmetric effect is also forbidden in
any planar structure containing a 4-fold axis of rotation and is
therefore irrelevant to the recent observation of the polarization
rotation in four-fold gammadion arrays \cite{Kuwata}.

Our theoretical analysis shows that in lossy anisotropic planar
chiral structures the asymmetry of total transmission is accompanied
by similar asymmetries in reflection and absorption. Nevertheless,
the total energy is always conserved. The spectral dispersions of
transmission, reflection and absorbtion asymmetries are, in general,
different and can therefore take zero values at certain frequencies.
For instance, at frequency $4.89 GHz$ absorption is the same for both
directions of propagation. Here the asymmetric transmission is due
solely to the asymmetric reflection. At $16.3 GHz$ transmission
asymmetry vanishes, but the asymmetry of reflection exists and is due
to the asymmetry of absorption.

\begin{figure}
\includegraphics[width=80mm]{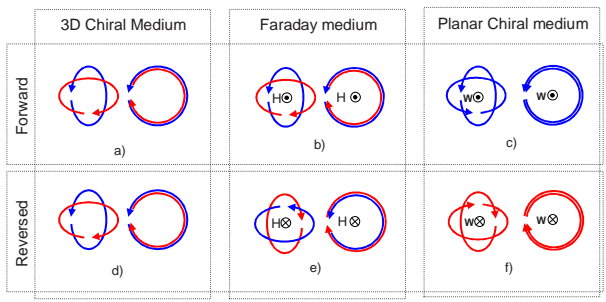}
\caption{Polarization eigenstates in an anisotropic three-dimensional
chiral medium (a left); an isotropic three-dimensional chiral medium
(a right); anisotropic Faraday medium (b, left); an isotropic Faraday
medium (b, right); an anisotropic planar chiral medium (c, left). The
single asymptotic polarization eigenstate for a planar chiral
structure with vanishing anisotropy is shown in (c, right). Parts
(e), (e) and (f) show the corresponding polarization eigenstates for
an electromagnetic wave propagating in the opposite direction to that
of cases (a), (b) and (c), correspondingly.} \label{eigen}
\end{figure}

Knowledge of the complex transmission matrix enables the calculation
of the polarization eigenstates of the system, i.e. the polarization
states that are not affected by transmission through the planar
structure  (see Fig 5). In analog to the anisotropy of bulk media,
the eigenstates of an anisotropic non-chiral planar structure are two
mutually perpendicular linear polarizations. In a loss-less chiral
structure they are also linear. However, the eigenstates become
elliptically polarized in lossy planar chiral structures. In contrast
to the Faraday effect, or conventional three-dimensional chirality in
bulk media for that matter, where the eigenstates are a pair of
\emph{counter-rotating} elliptical states, the eigenstates of a lossy
planar chiral structure are two \emph{co-rotating} elliptical
polarizations as illustrated in Fig.~5. These eigenstates only differ
in the azimuths of their main axes (they are orthogonal). As in the
case of the Faraday effect (but in contrast to gyrotropy in
conventional three-dimensional chiral media), the sense of rotation
of the elliptically polarized eigenstates is reversed for the
opposite direction of propagation through a planar chiral structure.
It is interesting to note here that if a planar chiral structure is
somehow continuously transformed so as to retain its geometrical
chirality but to reduce its anisotropy, the eigenstates will tend
towards a degenerate single circularly polarized eigenstate. This
eigenstate has only asymptotic meaning because in isotropic planar
chiral media the polarization conversion effect vanishes.

We derived the ellipticity of the eigenstates from the eigenvectors
of the experimental transmission matrix $t_{ij}$ and also calculated
corresponding theoretical values for the matrix obtained by the
method of moments. In both cases the frequency dispersion of the
ellipticity strongly resembles the dispersion of $\Delta \Xi$ with
the maximum ellipticity angle reaching 35 deg. As expected, the
polarization eigenstate ellipticities have opposite signs for
counter-propagating directions and for enantiomeric samples. Here the
ellipticity $\eta$ of the polarization state is defined in standard
fashion as $\eta = 0.5 \arctan(S_3 / S_0)$, where $S_3$ and $S_0$ are
components of the the Stokes Parameter of the polarization state.

In conclusion, circular conversion dichroism in a planar chiral
structure is a unique, previously unknown effect, which results from
the chirality {and anisotropy} of the structure and is inherently
linked to dissipation in the substrate. An attempt to describe it in
terms of an effective medium approximation, and in particular to
attribute it to the standard classification scheme of local and
non-local reciprocal and non-reciprocal polarization phenomena in
homogeniouse media \cite{Smolenskii,Birrs}, fails and is not
appropriate. Therefore this phenomenon is perhaps more closely
related to but not the same as the theoretically predicted phenomena
of non-reciprocal transmission through the interface between local
and non-local media \cite{Svirko}, and circular differential
reflectivity of planar chiral interface \cite{Bedeaux}. We expect
that the asymmetry of transmission through an appropriately scaled
sub-wavelength chiral planar structure could be seen in optical
experiments.

\begin{acknowledgments}
The authors would like to acknowledge the financial support of the
Engineering and Physical Sciences Research Council, UK and the EU
Network of Excellence `Metamorphose'.
\end{acknowledgments}

\bibliography{Chiral_FS}

\end{document}